# Puzzling behavior of hydrated YBCO

A. V. Fetisov

*Institute of Metallurgy of the Ural Branch of the Russian Academy of Sciences, Ekaterinburg, Russia, fetisovav@mail.ru*

**Abstract** We study the influence of a close-lying non-magnetic and non-charged metal disk on the hydration reaction of $YBa_2Cu_3O_{6+\delta}$ (YBCO). It has been established, that at a distance in the range of 6 – 7 mm from the disk surface, the rate of the hydration is significantly reduced relative to distances out of this interval. The effect obtained is very puzzling. It allows assuming an interaction of unknown nature to exist between YBCO and the disk. Further direct measurements of the force acting between these bodies have been performed. A recorded repulsive interaction with a magnitude of $6.5 \cdot 10^{-6}$ N, which is 0.04% of the sample weight, has confirmed our assumption.

## 1 Introduction

Recently, there were reports that the $^8$Be and $^4$He isotopes, serving as objects of studies at the subatomic level, exhibit unique anomalous properties concerning angular correlations of electron-positron pairs created inside them. In order to interpret their behavior, the authors were forced to assume that observed anomalies could be associated with the birth and subsequent decay of a certain neutral particle – boson with mass $m_0c^2$ = 16.70 MeV [1, 2]. On the basis of that finding, the authors of [3, 4] supposed that the found hypothetical boson is a new vector gauge boson that mediates a fifth force of nature. Analyzing the results of our work, we believe that we are faced with an equally extraordinary phenomenon. The $YBa_2Cu_3O_{6+\delta}$ (YBCO) superconductor hydrated in a special way at low $p_{H_2O}$ and room temperature (RT) behaves very exotically, remotely interacting with surrounding bodies which exhibit neither magnetic nor electrical properties [5, 6]. This can be indirectly observed, for example, by changes in the rate of the hydration of YBCO near a massive metal disk [6]. It is surprising that interaction with the disk occurs in a narrow range of distances $l$ from it with maximum intensity at $l$ = 6–7 mm. It could be assumed that there is a certain field affecting the hydration reaction (in [6] it was assumed that this field is electromagnetic). At the same time we have so far failed to record anything emanating from YBCO under the conditions of our experiment, with the exception of a very weak field from electrostatic charge (which will be discussed in the framework of this paper). This charge quickly accumulates on the samples but, however, and quickly disappears at the very beginning of the experiment. Our research echoes the works of E. Podkletnov et al. [7, 8], in which a partial weight loss of samples of different material and different chemical composition suspended over a massive superconducting

YBCO disk was recorded. Although the authors of [7, 8] explained their results from their special perspective, it is likely that we observed the same phenomenon.

The main aim of the present study is to record the interaction found in [6] by the "direct" method – as the force acting from the side of YBCO on a metal body. For this purpose, we isolated YBCO samples from the external environment, thereby excluding all possible mass transfer processes between them, and then controlled change in the weight of samples together with an insulating system. The created closed system had all necessary conditions for YBCO to be hydrated at low $p_{H_2O}$. The fact of changing the weight of the insulating system with samples during hydration was supposed to prove the existence of the force interaction between YBCO and surrounding bodies, which cannot be explained in terms of the concept of the four interactions.

## 2 Experimental Details

The YBCO material for this study was prepared by the method described in [5, 6]. The resulting YBCO powder was poured into quartz cups of 7.3 mm diameter and 13 mm length. A mass of powder portions varied around 1.7 g. Cups with YBCO were covered with two layers of aluminum foil of 11 μm thickness; several small holes were made in the foil to allow air to come into the cups. Aluminum foil provided shielding of the samples from external electric fields. Prepared samples were exposed to air with 3.0 % relative humidity for 1 – 15 days. Then each cup with YBCO was placed into a sealed glass bottle with moisture holding material providing a humidity of 43% ($K_2CO_3 \cdot 1.5H_2O$).

A study of the change in weight with time of the bottles with samples were carried out using an analytical balance Shimadzu AUW-120D (Japan). The weight changes were measured relative to identical bottles without samples. Near the location of the bottles there was placed a barometer that made it possible to check the gas tightness of the bottles with samples. If the gas tightness was broken, then significant weight fluctuations were observed over time, which were synchronous with changes in atmospheric pressure.

In the course of studying it was noticed that small amount of electrostatic charge accumulating on the samples during experiments can affect sample properties. In order to further control the charge accumulation, further experiments were carried out using three types of bottles, which differ in the efficiency of the electrical connection of the samples with environment. So, bottle *1* remained the original – in it a sample did not come into contact with anything except the bottom of the bottle. Bottles *2* and *3* allowed the sample to be in close contact with the cap, which had relatively good conductivity. The outer surface of bottle *2* also had a ground loop. The accumulation of charge on the samples was recorded by measuring the force of their attraction to a metal disk located at a distance of 5 mm.

The X-ray diffraction analysis (XRD) was performed using a Shimadzu XRD-7000 diffractometer (Cu$K_\alpha$-radiation, Bragg angle range 2Θ = 20 ÷ 80°, a step of 0.02°, counting time of 2 s per step). Crystal structure analysis was performed using GSAS package [9] starting from the model of the crystal structure presented in [10]. The follow-ing discrepancy factors were achieved: weighted profile $\omega R_p \approx 8$–$10\%$, unweighted structural $R_f = 5$–$8\%$, Durbin-Watson statistic 0.5–0.9.

## 3 Experimental Results

### 3.1 X-ray diffraction and microscopy

X-ray diffraction experimental and calculated data obtained on the initial and hydrated samples are shown in Fig. 1. YBCO samples are crystallized in the orthorhombic structure (*O*) with the *Pmmm* space group. Their unit cell parameter values are presented in Table 1. Fig. 2 represents a more detailed fragment of the spectrum in Fig. 1 which includes the characteristic peaks of impurities usually detected in YBCO. Here we can see a small peak of $Y_2BaCuO_5$ at about $2\Theta = 30°$ which is already present in the initial compound. For hydrated material one can observe besides such impurities as $Ba_2Cu_3O_5$ and "$Ba_4Cu_2O_{12}$",[1] which have formed as a result of interaction YBCO with water at low $p_{H_2O}$. Meanwhile, the production of impurity phases is concentrated at the beginning of the hydration process (for the first three days) and does not occur after that. At this, the degree of saturation of YBCO with water, during the interval from 3 to 14 days, has grown from 0.27 to 0.65 wt.%. This is consistent with the results of our earlier work [5], which have shown that the saturation of YBCO with water at low $p_{H_2O}$ occurs mainly without the formation of strong chemical bonds.

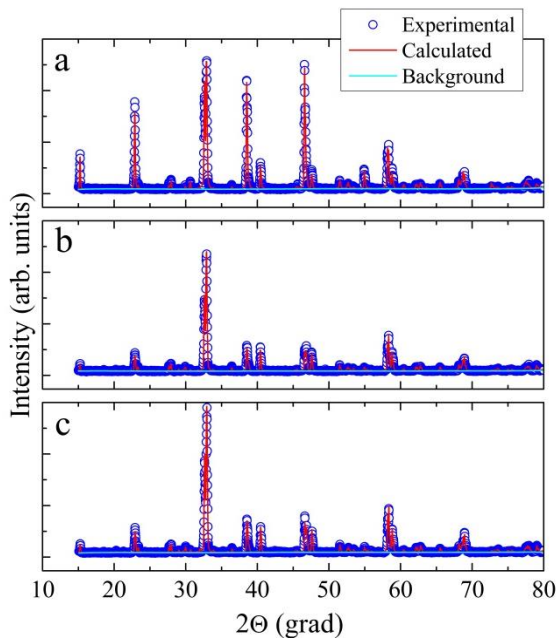
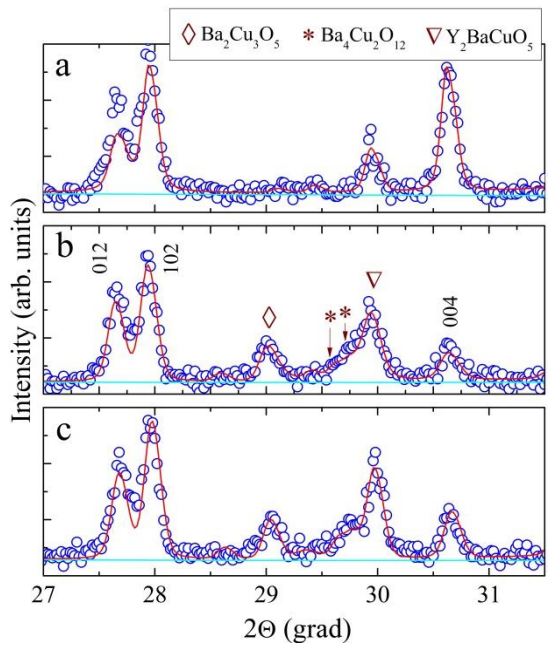

**Fig. 1** X-ray diffraction pattern of the YBCO samples: initial (a) and hydrated for 3 (b) and 14 days (c).

**Fig. 2** Region of the X-ray diffraction pattern of YBCO (Fig. 2) with main reflects of the impurities contained in the samples.

---

[1] Their models of the crystal structure are presented in [11, 12].

**Table 1** The structural data and phase composition of the YBCO samples before and after exposure to humid air

| Structural parameter | Before exposure to humid air | After exposure to humid air for 3 days | After exposure to humid air for 14 days |
|---|---|---|---|
| Major phase: orthorhombic, space group *Pmmm* | | | |
| $a$ (Å) | 3.82699 | 3.82540 | 3.82456 |
| $b$ (Å) | 3.88460 | 3.88216 | 3.88271 |
| $c$ (Å) | 11.71196 | 11.70221 | 11.70444 |
| $V$ (Å$^3$) | 174.114 | 173.787 | 173.807 |
| Impurities (Wt.%) | $Y_2BaCuO_5$ (2.2) | $Y_2BaCuO_5$ (3.8) $Ba_2Cu_3O_5$ (1.0) $Ba_4Cu_2O_{12}$ (1.9) | $Y_2BaCuO_5$ (3.9) $Ba_2Cu_3O_5$ (0.8) $Ba_4Cu_2O_{12}$ (1.7) |

Fig. 3 shows the images of the initial and hydrated YBCO powders obtained using an optical microscope at a small magnification. It can be seen that in the course of the hydration process proceeding at low $p_{H_2O}$ and resulting in a small amount of impurities, the YBCO material nevertheless easily agglomerates. It has been noticed that the more time the hydration at low $p_{H_2O}$ flows, the more strength the agglomerated YBCO acquires. Meanwhile, other YBCO powders have been in our laboratory for several decades under natural humidity (up to 2.4 kPa) and they have greatly degraded. However, its flowability has hardly changed. So, in fact, caking processes at RT are not characteristic of YBCO material. All this suggests a possible additional interaction between YBCO particles when hydrating at low $p_{H_2O}$.

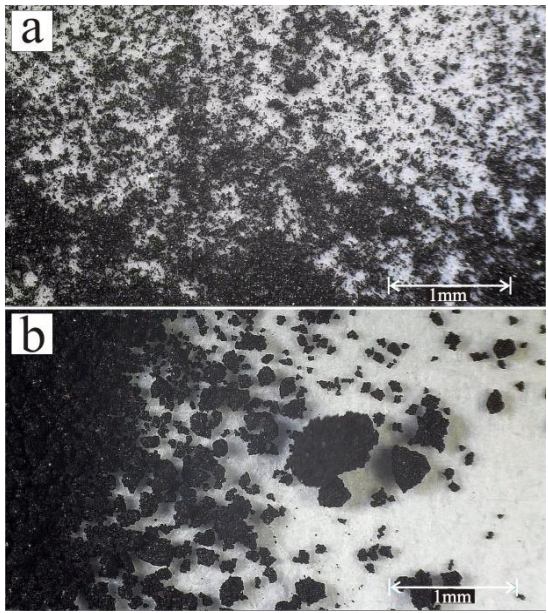

**Fig. 3** Images of the slightly ground samples: initial (a) and after 14 days hydration (b).

## 3.2 Investigation of new specific properties of YBCO arising during its hydration

The regimes of the heat treatment of samples, while they were being exposed in the atmosphere with 3.0 % relative humidity on the final stage of YBCO preparation, are listed in Table 2 (col. 2). According to our preliminary studies [5, 6], this stage is the most important for the formation of specific properties of the material. Also listed in Table 2 are these properties themselves and their numerical parameters corresponding to different batches of samples. In the order of listing, the first parameter is: the change in the hydration rate of YBCO occurring under the influence of metallic body – the effect studied in detail in [6] (hereinafter referred to as *effect I*). It was found that a sample located ~7 mm from the surface of a massive metal disk (DN40CF flange) undergoes the hydration process to a much lesser extent than one located far away from the disk. In the present study, ratios of hydration rates for such pairs of samples (for each batch we used two pairs at the same time and the result was averaged) were obtained and given in Table 2 (col. 3). *Effect I* was supposed to be used as a control – it should show how close the regimes of heat treatment used in the work were to "optimal", suitable for creating the material with a spectrum of specific properties. New properties studied in this work were: the rate of change in the weight of the sealed bottles with samples (hereinafter referred to as *effect II*, col. 4) and, in addition, the presence of the charge accumulated on the samples as a result of processes occurring in the bottles (col. 5).

**Table 2** Regimes of heat treatment of samples and a set of quantity and quality parameters obtained in experiments

| No. of batch | $T$ ($^0$C) and $\tau$ (days) [*] | Change in hydration rate (%) | Average rate of weight change (mg/day) | Appearance of charge | No. of bottle | No. of batch | $T$ ($^0$C) and $\tau$ (days) | Change in hydration rate (%) | Average rate of weight change (mg/day) | Appearance of charge | No. of bottle |
|---|---|---|---|---|---|---|---|---|---|---|---|
| 1 | 2 | 3 | 4 | 5 | 6 | 1 | 2 | 3 | 4 | 5 | 6 |
| B.693 | 27.1/2 | -28.0 | -0.0200 | - | 2 | B.710 | 25.5/13 | 10.0 | -0.0050 | Yes | 1 |
| B.694 | 26.9/3 | -52.5 | -0.0060 | - | 3 | B.711 | 25.5/15 | -14.0 | -0.0930 | Yes | 2 |
| B.695 | 26.9/1 | -9.0 | -0.0008 | - | 1 | | +25.9/1 | | | | |
| B.696 | 26.9/3 | -26.0 | -0.0200 | - | 2 | B.712 | 25.5/4 | -4.0 | -0.0030 | Yes | 1 |
| B.697 | 25.2/4 | -12.0 | 0.0042 | - | 3 | B.713 | 25.5/5 | -9.2 | 0.0040 | Yes | 3 |
| B.699 | 25.5/5 | -55.3 | 0.0006 | - | 2 | B.714 | 25.5/6 | -28.5 | -0.0076 | No | 2 |
| B.700 | 25.4/5 | -45.0 | -0.0088 | - | 3 | B.715 | 25.5/7 | 2.8 | -0.0010 | Yes | 1 |
| B.701 | 25.2/11 | -0.1 | 0.0030 | - | 1 | B.716 | 25.5/9 | 7.5 | 0.0020 | Yes | 3 |
| B.702 | 25.2/13 | -4.3 | -0.0010 | - | 2 | B.717 | 25.5/11 | 3.8 | -0.0114 | No | 2 |
| B.703 | 25.2/14 | -12.0 | 0.0030 | - | 3 | B.718 | 25.5/1 | -1.0 | -0.0012 | Yes | 1 |
| B.704 | 23.0/15 | -22.4 | -0.0095 | No | 2 | B.719 | 22.7/6 | 7.2 | 0.0015 | Yes | 3 |
| B.707 | 25.5/10 | -2.0 | 0.0025 | - | 2 | B.720 | 25.5/2 | -11.3 | 0.0010 | Yes | 3 |
| B.709 | 25.5/12 | 2.1 | 0.0010 | Yes | 3 | [*]Temperature/Time | | | | | |

It follows from Table 2 that the regimes of preliminary processing of samples, indeed, play a crucial role in the formation of their future properties. This is most clearly manifested in slowing down the kinetics of the hydration process near a metal body (*effect I*). This effect is sharply enhanced near the regime parameters: $T = 25.5$ °C; $\tau = 5 - 6$ days. An increase in temperature to 27°C and a decrease to 23°C result in new optimal values for the parameter $\tau$: 2 – 3 days and 15 days, respectively.

Fig. 4 shows examples of kinetic curves obtained on bottles with various samples (hereinafter, samples are designated as their batch numbers). Since the mass transfer of bottles with the surrounding space has been technically excluded, it is assumed that the change in the weight of bottles is a manifestation of specific properties of YBCO and occurs due to the interaction of the YBCO samples with surrounding bodies. In this case, it is probably a massive metal base of the balance located at a distance of 15 mm from a sample.

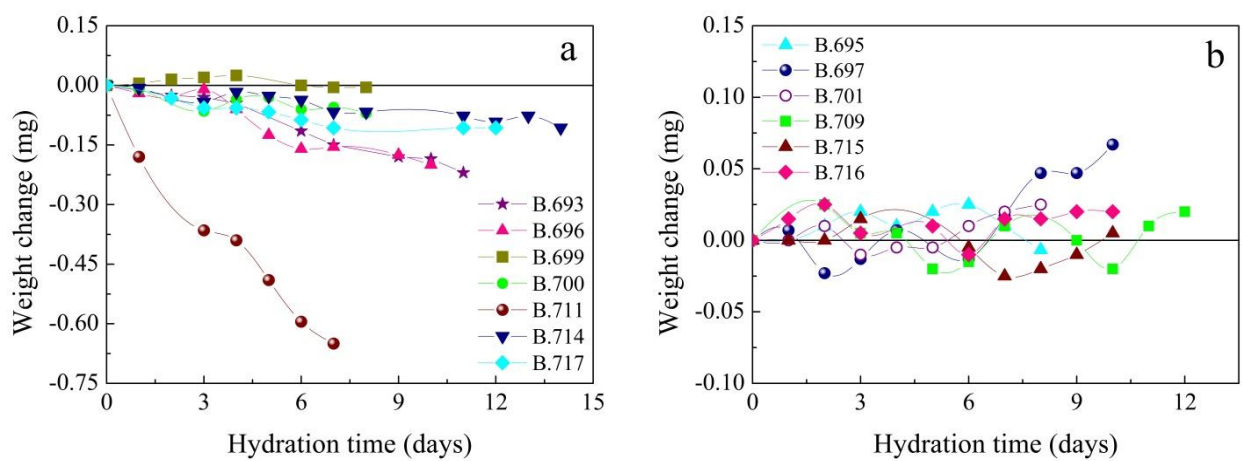

**Fig. 4** Weight change kinetics of some samples (**a**) demonstrating the effect of hydration rate reducing under the influence of metal vicinity; (**b**) not demonstrating that.

A review of kinetic features of the weighed samples (see Table 2 and Figure 4) shows a uniform decrease in the weight of samples demonstrating a moderate negative *effect I*. In turn, samples that do not exhibit *effect I*, or conversely if this effect is significant keep their weight unchanged (with very rare exceptions). Fig. 5 clearly shows how changes in the weight of samples are related to the magnitude of *effect I*. Meanwhile, as it turns out, vivid manifestations of *effect II* can also be associated with the fact that, by chance, almost all the corresponding experiments have been carried out using bottle *2*, in which a quick removal of electrostatic charge from a sample has been ensured.

Kinetic curves of charge acquired on the samples during their exposure in sealed bottles are depicted in Fig. 6. Shown on the right axis of Fig. 5 graph is correspondence of an accumulated charge value to the magnitude of *effect I*. Here, the arbitrary unit of charge approximately corresponds to $3 \cdot 10^{-11}$ C.

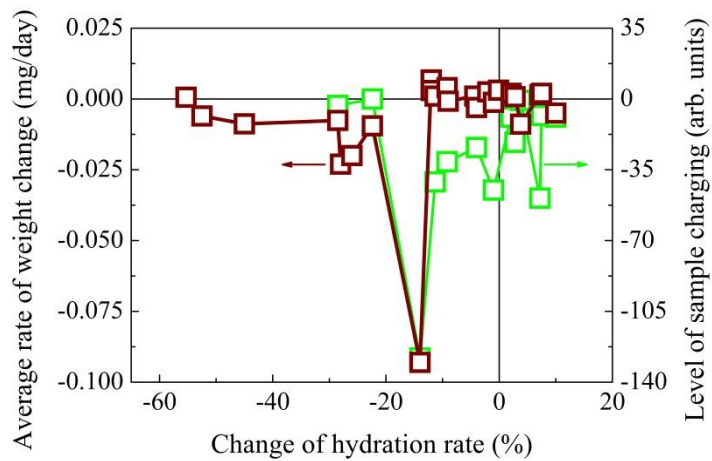

**Fig. 5** Matching of the ability of some YBCO samples to change in weight (*effect II*, *left axis*) with the change of YBCO hydration rate occurring under the influence of metal vicinity (*effect I, bottom axis*), see data in Table 2. There is also represented the charge acquired by YBCO during the first day of hydration (*right axis*).

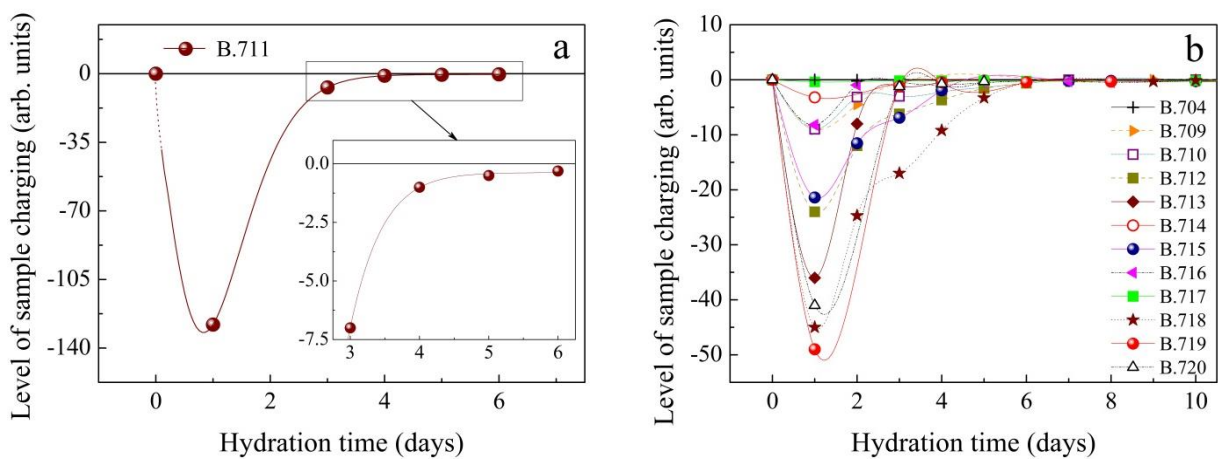

**Fig. 6** Electrostatic charge acquired by YBCO during hydration.

The extreme character of the dependencies in Fig. 6 is explained by the fact that nature of the charge accumulation is likely related to the change in the weight of samples, and the charge accumulation intensifies with a sharp initial increase in weight. If the charge does not have time to dissipate into the surrounding space, then a further change in weight, according to the Le Chatelier's principle, may be blocked. At least, from these positions the course of the dependences in Fig. 6 can be easily explained taking into account the fact that the charge dissipation rate should increase in the order: bottle *1* → bottle *3* → bottle *2* (see paragraph 2). For example, the charge on bottle *1* (see samples B.712, B.715, and B.718), having reached its extreme value by the first day, subsequently does not quickly disappear. Then, according to the above, it should block the development of processes responsible for the change in the weight of samples in this bottle. As a result, not a single sample from bottle *1* has shown a noticeable change in weight. On the other hand, most of the samples located in bottle *2* have shown significant changes in weight (see Table 2), which may be related exactly to the possibility of rapid dissipation of the accumulated charge. Indeed, as can be seen from Fig. 6, the charge on this bottle (see samples B.704, B.714, and B.717) is either very weak or not observed at all.

Thus, we have two possible explanations for why some samples show *effect II*, while others do not. And it seems to us that in fact both factors can play a role here: the correspondence to the optimal preparing parameters before an experiment and the possibility of rapid charge dissipation during hydration.

In conclusion, we would like to discuss the fact that in order to intensively reduce the weight of sample B.711, the simplest alternative explanation is clearly begged: the poor gas tightness of the bottle in which the sample has been located during the experiment. Meanwhile, to control the gas tightness we used a comparative analysis of the course of changes in the weight of bottles with the course of changes in atmospheric pressure, as mentioned above. It is important to emphasize that the change in the weight of test bottles was determined relative to the change in the weight of tightly sealed identical bottles, which are in equal conditions but without samples. As it is seen from Fig. 7, the course of dependences "Weight change vs. hydration time" for the case of leaky bottles almost exactly repeats the course of changes in atmospheric pressure. Moreover, the behavior of these dependences is close to the calculated from the equation:

$$Weight\ change \approx \rho_0 \cdot V \cdot (\Delta P/P_0), \quad (1)$$

where $\rho_0$ is the air density at the atmospheric pressure: $(P_0 + \Delta P)/2$ and the temperature of the experiment (297 K); $V$ is the volume of bottles (15 ml); $\Delta P$ is the change of atmospheric pressure that has occurred since the beginning of the experiment; $P_0$ is the atmospheric pressure at the beginning of the experiment (about 100 kPa). Small deviations from the theoretical dependences are associated with the mass transfer between leaky bottles and the environment. At the same time, the course of the dependence of sample B.711 does not correlate with changes in atmospheric pressure that serves as an evidence of a sufficiently good tightness of the corresponding bottle[2]. As for the effect of electrostatic charge on the weight, as shown by special experiments, this effect is insignificant and opposite to the decreasing weight of most samples.

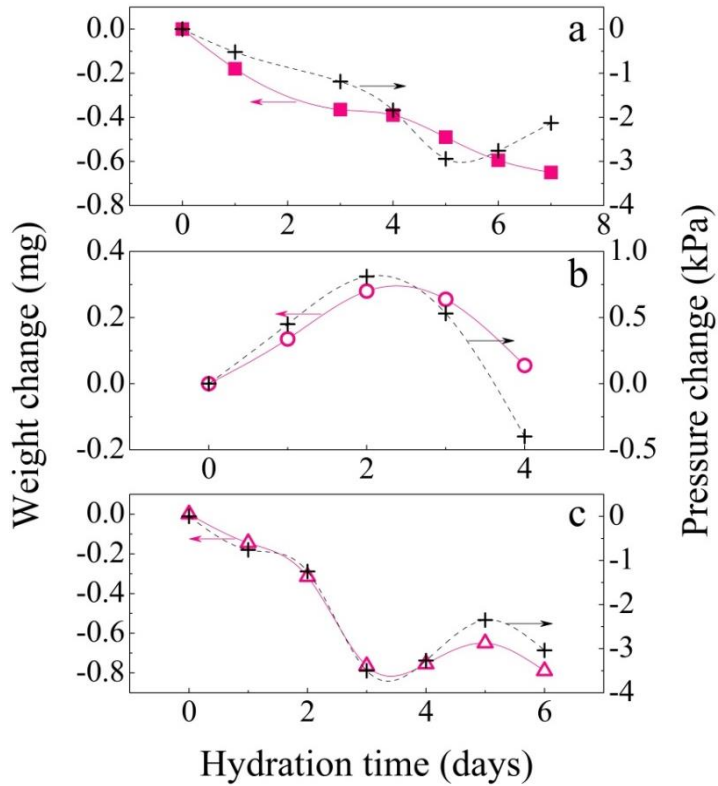

**Fig. 7** Matching of the changes in the weight of: sample B.711 (**a**), empty vessel (**b**), and sample B.708 (**c**) with the changes in atmospheric pressure occurred during the experiments. In the cases (**b**) and (**c**) there was bad sealing of the vessels.

## 4 Discussion

One of the goals of this study was to explain the experimental results obtained in [6] – the extreme dependence of the hydration rate of YBCO on the distance to the metal disk, which according to all known laws should be neutral with respect to this reaction. In the above study, a noticeable force interaction was observed between hydrated YBCO and a metallic body located in close proximity. It was also found that this interaction is accompanied by the accumulation of electrostatic charge on the YBCO material. On the basis of the new data we will try to discuss the extreme character of the dependence found in [6], without going into details of underlying mechanisms of *effect I*.

    We note that the rapid charge removal from the sample apparently stimulates not only the manifestation of *effect II*, but also *effect I* (see Fig. 5, data points obtained for the samples B.704 and B.714). If these data are taken into account, one can assume the removal of the charge to be one of the functions of a massive metal body during the implementation of *effect I*. In this case, the metal body plays the same role as the ground loop for bottle *2* (see paragraph 2). The closer the container with a sample to it, the faster flow of charge from the sample is and, this results in intensifying *effect I*. But, taking only this factor into account, one cannot obtain the extremal dependence observed in [6] with a maximum at $l = 6$–7 mm. However, this dependence can be obtained if we consider as a cause-and-effect chain the following: the removal of charge stimulates *effect II*, which, in turn, intensifies *effect I*. Nonmonotonicity in this case is hidden in the dependence "*effect I* vs. *effect II*", which can be obtained from the experimental data in Fig. 5.

    Fig. 8a presents the fitting dependence describing the experimental data "*effect II* vs. *effect I*" in Fig. 5 with using the asymptotic function:

$$y = -0.00129 \cdot x \cdot (x + 13.80655)^{-1}. \tag{2}$$

In Fig. 8b one possible solution of the inverse function $x = f(y)$ is shown. It describes the dependence "*effect I* vs. *effect II*" relating to the set of experimental data. Taking the parameter *l*, which is linked with *effect II*, as the upper horizontal axis, we thus have ob-tained a schematic dependence "*effect I* vs. distance *l*" (according to experiments with bottles *1–3* of different conductivity, parameter *l* monotonically weakens *effect II* as it increases). As can be seen from the figure, this dependence contains a sharp extremum, which is very similar to that obtained for this dependence in [6].

---

[2] Similar conclusions were made for all other samples of our study.

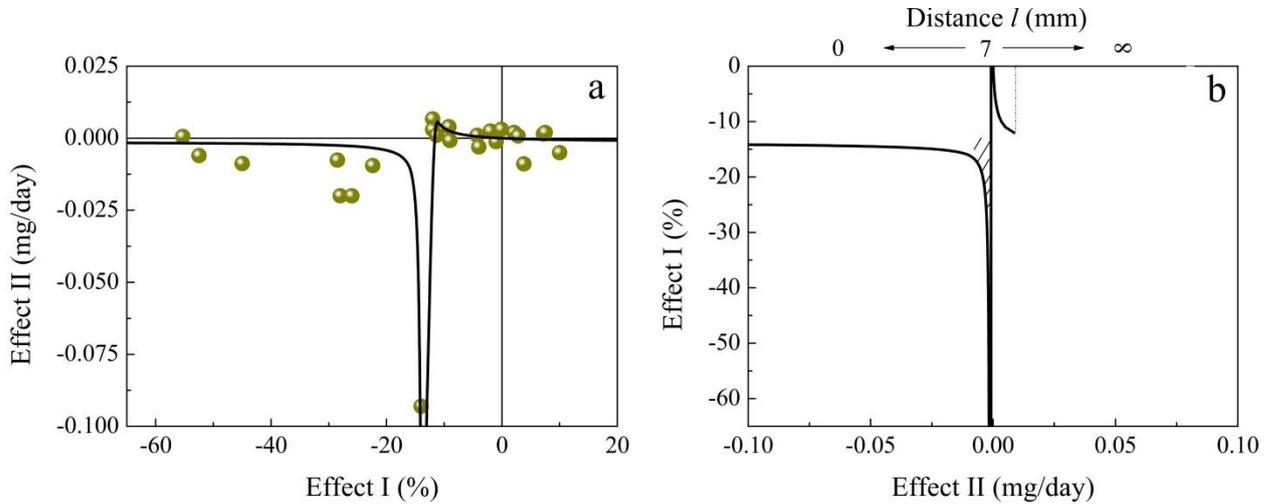

**Fig. 8** Direct and inverse relationships linking *effect I* and *effect II*.

Thus, the experimental data of the present work and the extraordinary data obtained in [6], as it turns out, are in good agreement, which in itself is an important result. In addition, here we come to some conclusions about the cause of the sharp extremum on the dependence "*effect I* vs. distance *l*". So, it is due to the extreme behavior of the hydration reaction of YBCO depending on the magnitude of a certain repulsive impact from YBCO to the surrounding matter. One can therefore assume that, in addition to metal bodies, water molecules are also exposed to this impact and, as a result, change their activity.

In conclusion, we will try to consider another important question: whether the massive metal body is involved in the inhibition of the hydration reaction in some other way, in addition to its role of the grounding contour? To answer that question, the data of some experiments in which *effect II* has been studied should be taken in consideration. These experiments have been carried out without the use of massive metal body, however, with using bottle *2* there has been provided a good charge removal too. Then the question posed above can be transformed into the following: was *effect I* realized in bottles *2* in the same way as it was realized in specialized experiments (according to the data of col. 3 in Table 2)? Since in the case of bottles 2 the samples themselves (without bottle) have been weighed only at the beginning and at the end of the experiment with different hydration time $\tau$, we are forced to compare the change in their weight with the corresponding point $\tau$ on a reference kinetic curve of hydration. Such the reference kinetic curve was obtained in a special experiment on a sample not subjected to the influence of nearby massive bodies. The obtained curve has been scaled for each batch so as to satisfy the change in weight $\Delta W$ of the sample not subjected to the influence of massive metal body in the experiment studying *effect I*. This weight change occurred within 4 days (duration of this type of experiment [6]), therefore, on the scaled reference curve the point (*x*; *y*): $\tau$ = 4 days; $\Delta W$ should lie.

Fig. 9 presents the comparative analysis of the hydration rate for the samples taking part in *effect I* and *effect II* without any obstacles for removing electrostatic charge. It shows that without massive metal body located near the sample the inhibitory effect for the hydration of YBCO is not manifested. In other words, massive metal body plays a non-trivial role in its implementation.

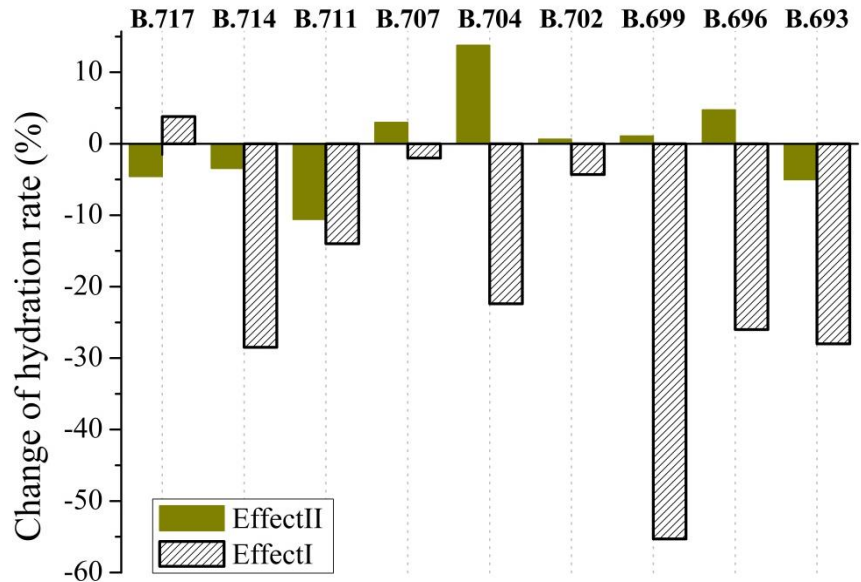

**Fig. 9** Implementation of *effect I* in the samples for which as an initial project was study of *effect II* (dark yellow strips) in comparison with the samples for which study of *effect I* was as an initial project (black-white strips).

It should be noted that the hydration deceleration observed for sample B.711 in Fig. 9 (series "Effect II") is imaginary because the corresponding decrease in the sample weight has occurred in good agreement with the kinetic curve in Fig. 4a. This curve is assumed to correspond to an unknown force interaction of the sample with surrounding bodies. At the same time, we cannot yet explain the enhanced hydration of sample B.704.

## 5 Conclusion

Thus, the use of a specific procedure in the preparation of YBCO has allowed obtaining material with new properties:
- hydration of this material begins to proceed at such low $p_{H_2O}$ values as 1.1–1.3 kPa;
- it is observed a repulsive interaction of YBCO with surrounding metal bodies of $6.5 \cdot 10^{-6}$ N. It is possible that the repulsive interaction also occurs between YBCO and water molecules;

- At the beginning stage, the hydration of YBCO is accompanied by insignificant electrostatic charge accumulation on the superconductor. This process prevents the appearance of the repulsive interaction. So, in order to this interaction to be realized, it should be used any methods of removing the charge from the sample.

The discovered new properties of hydrated YBCO can be attributed to the manifestation in this compound of an interaction of unknown nature. In this connection, it is logical also to assume that this new interaction is somehow involved in the appearance of high-temperature superconductivity in YBCO oxide.

**Acknowledgments** The author thanks E.A. Vyaznikova for providing X-ray diffraction data. The study was performed with the use of equipment of the Ural-M Colletive Use Center at the Institute of Metallurgy of the Russian Academy of Sciences.